\documentclass[]{aa}
\usepackage{natbib}
\usepackage{lscape,graphicx} 
\usepackage{epsf}
\usepackage{subfig}
\usepackage{txfonts}
\normalfont

\begin{document}

\title{Infrared photometry of Cepheids in the LMC clusters NGC~1866 and NGC~2031\thanks{Based on observations collected at the Las Campanas Observatory of the Carnegie Institution of Washington, and at the European Southern Observatory, La Silla, Chile, using SOFI at the 3.5m NTT, within observing program 68.D-0287.} } 

\author{Vincenzo Testa \inst{1} \and Marcella Marconi \inst{2} \and Ilaria Musella \inst{2} \and Vincenzo Ripepi \inst{2}  
\and Massimo Dall'Ora \inst{2} \and Francesco R. Ferraro \inst{3} \and Alessio Mucciarelli \inst{3} \and Mario Mateo \inst{4}
\and Patrick C\^ot\'e \inst{5}}

\institute{INAF - Osservatorio Astronomico di Roma, Via Frascati 33, 00040 Monte Porzio Catone, ITALY \and
INAF - Osservatorio Astronomico di Capodimonte, Via Moiariello 16, 80131 Napoli, ITALY \and
Dipartimento di Astronomia, Univ. di Bologna, Via Ranzani 1, 40127 Bologna, ITALY \and
Department of Astronomy, University of Michigan, Ann Arbor, MI 48109-1090, USA \and
Herzberg Institute of Astrophysics, National Research Council of Canada, 5071 West Saanich Road, Victoria, BC, V9E 2E7, Canada}

\offprints{V. Testa, testa@oa-roma.inaf.it}

\date{Received Juky 26, 2006/ Accepted October, 11 2006}

\abstract
{Near infrared (IR) studies of Cepheid variables in the LMC take advantage of the reduced light curve amplitude and metallicity dependence at these wavelengths. This work presents such photometry for two young clusters known to contain sizeable
Cepheid populations: NGC 1866 and NGC 2031.}
{Our goal is to determine light curves and period-luminosity (PL) relations in the near-IR,  to assess the similarity between cluster and field pulsators, and to examine the predictive capability of current pulsation models.}
{The light curves are obtained from multiwavelength broadband J,H,K$_S$ photometry of Cepheids in both clusters, with periods previously established from optical photometry.}
{Mean magnitudes for the Cepheids are used to construct PL relations in the near-IR.  The properties in the PL planes are compared with the behavior of field Cepheids in the LMC and with the predictions of recent pulsational models, both canonical and overluminous.}
{Cluster and field Cepheids are homogeneous and the inclusion of the cluster Cepheids in the field sample extends nicely the PL relation. The slope of the PL relation is constant over the whole period range and does not show -- at least in the adopted IR bands -- the break in slope at $\mathrm{P \sim 10~d}$ reported by some authors. A comparison with the predictions of pulsation models allows an estimate for the distance moduli of NGC~1866 and NGC~2031. The two clusters are found to lie at essentially the same distance. Fitting of theoretical models to the data gives, for the K filter, (m-M)$_0$ = 18.62$\pm$0.10~mag if canonical models are used and (m-M)$_0$ = 18.42$\pm$0.10~mag if overluminous models are used. On the basis of this result, some considerations on the relationship between the clusters and the internal structure of the LMC are presented.}

\keywords{Cepheids - Magellanic Clouds - globular clusters: individual: NGC 1866 - globular clusters: individual: NGC 2031 - Stars: distances}

\authorrunning{Testa et al.}
\titlerunning{Infrared Photometry of Cepheids in LMC Clusters}

\maketitle

\section{Introduction}

Cepheid variables are one of the cornerstones upon which the extragalactic distance scale is based. Since the pioneering work by Henrietta Leavitt in 1912 (Harvard College Observatory, Circ. No. 173), their importance has grown to the point that they are considered one of the most important distance indicators for galaxies within, and beyond, the Local Group. Recent HST studies have made use of Cepheids to measure the distances of galaxies beyond the Local Group \citep[e.g. the Extragalactic Cepheid project][]{Freedman01} and to thereby calibrate a variety of secondary distance indicators. In this regard, the Magellanic Clouds (MCs) --- and, in particular, the Large Magellanic Cloud (LMC) --- offer a unique laboratory for testing the Cepheid distance scale, being among the closest extragalactic environments with a resolved stellar population. Since LMC Cepheids lie very nearly at the same distance from the Sun,
they offer an attractive way of sidestepping the problem of relative distance errors that plague studies of the Period-Luminosity (PL) relation slope using Galactic samples. For these reasons, the LMC has long been an important first step in establishing the Cepheid distance ladder: i.e., the calibration of the period-luminosity (PL) relation in distant galaxies often relies on an assumed LMC distance modulus (DM). For instance, in the Extragalactic Cepheid project, a DM of $\mathrm{(m-M)_0 = 18.50}$~mag was adopted for the LMC \cite{Freedman01}.

Among the most pressing issues to be considered when dealing with the Cepheid PL relation is its dependence on the metallicity, which has been questioned in a number of past works. From an observational perspective, some authors \citep[e.g.][]{Kennicutt98,Sakai04,Tammann03, Romaniello05} argue that such a dependence is indeed present, while others \citep[e.g.][]{Gieren05} suggest that any effect is
small, perhaps ill-defined \citep[e.g.][]{Groenewegen03,Storm05}. 
On the theoretical side, some authors again find a mild dependence (if any) from linear nonadiabatic pulsation models \citep[see e.g.][]{Saio98,Alibert99}, whereas nonlinear convective pulsation models predict a significant dependence on both metal content and helium abundance \citep[see][ and references therein]{Fiorentino02,Marconi05}.

A determination of the PL relation presents different challenges in different wavelength intervals. In the optical bandpasses, the relation shows non-negligible scatter 
and the use of a color is required to reduce the dispersion (i.e., a Period-Luminosity-Color relation, or PLC). 
Moreover, the light curve amplitude is large so the measurement of the mean magnitude, on which both the PL and PLC relations are based, requires accurate photometry and well sampled light curves (although the large amplitude allows a better determination of the period because it is less sensitive to small scatter in the measurements). By  contrast, 
the light curve amplitude is much smaller at near-IR wavelengths and the determination of the mean magnitude correspondingly easier. Moreover, the dependence on  both the reddening and the color extension of the instability strip is strongly reduced leading to a narrower PL relation than in the optical case. On the other hand, the reduced pulsation amplitude makes the period determination more difficult. In this work, in fact, we make use of periods already available from a variety of literature sources. 

Theoretical predictions based on nonlinear convective models confirm the great advantage of using near-IR filters to construct PL relations. As the wavelength increases from optical to near-IR bands, PL relations become linear over the full period range spanned by the observed pulsators and their dependence on metallicity and the instrinsic width of the instability strip decreases significantly \citep[see, e.g.][]{Bono99,Caputo00}.

In this context,  classical Cepheids belonging to young stellar clusters  play a particularly relevant role. In fact, not only they are at the same distance but are also characterized by the same age and chemical composition, thus offering a unique opportunity to investigate the uncertainties affecting both empirical approaches and theoretical scenarios. For the present study, we selected two young clusters having the largest known Cepheid population among LMC clusters: NGC~1866 and NGC~2031. 

The first is the most massive young cluster in the age range $\sim$ 100--200 Myr. It has been  the subject of a very large number of papers in the past, beginning with those by \cite{Arp67} and \cite{Robertson74}. Subsequent authors focused on studying the cluster either as a testbed of stellar evolution theory \citep[e.g.][]{Brocato89,Chiosi89,Brocato94,Testa99,Walker01,Barmina02,Brocato03}, as a Cepheid host \citep{Welch91,Welch93,Gieren94,Walker95,Gieren00,Storm05}, or as a dynamical laboratory \citep{Fischer92}. It has also been the subject of a strong debate over the presence of convective overshooting in intermediate-age stellar models, and on the fraction of binaries on the main sequence. Because the cluster lies  in the outskirts of the LMC, field contamination is not severe.
NGC~2031, on the other hand, is an order of magnitude smaller in mass and is located in the LMC bar, resulting in a high contamination of its stellar population. As an added complication, the surrounding field has an age comparable to the cluster (as is apparent in the CMD shown in Fig. \ref{fig:cmd}). Despite its large Cepheid population, it has not been extensively studied in the past, except for an analysis of the CMD by \cite{Mould93} who quote an age of $\sim$140 Myr, very close to that of NGC~1866. 

The metallicity of NGC~1866 has been estimated by various investigators, with the spectroscopic determination of \cite{Hill00}, $\mathrm{[Fe/H] = -0.5 \pm 0.1}$ confirming the photometric determination of \cite{Hilker95}, $\mathrm{[Fe/H] = -0.46 \pm 0.18}$. The first --- and, to our knowledge, only --- spectroscopic determination of the metallicity of 
NGC~2031 by \cite{Dirsch00} yielded $\mathrm{[Fe/H] = -0.52 \pm 0.21}$, indistinguishable from NGC~1866.
Moreover, both clusters have a fairly populated AGB that is clearly defined as a narrow sequence in the near-IR CMD (see Fig. \ref{fig:cmd}), and the
most luminous stars have been subject of a spectroscopic study \citep{Maceroni02} that confirms the evolutionary status of thermally-pulsing AGB stars at the tip of the AGB sequence.
Taken together, these properties suggest that NGC~2031 can be considered a low-mass analogue of NGC~1866.

The paper is structured as follows: the observations and reductions are described in Sect. \ref{sect:obs}; Sect. \ref{sect:phot} deals with the presentations of the photometric sample of the Cepheids; in Sect. \ref{sect:pl}, IR light-curves and PL relations are presented along with a comparison to previous  results and to the predictions of theoretical models; we conclude with a summary of our results in Sect. \ref{sect:summ}.

\section{Observations, Reductions and Calibrations \label{sect:obs} }

\subsection{Observations}
Data were collected over four observing runs spanning an interval of nine years. Although the phase sampling is sparse and quite random, we show later that is not a serious issue in the near-IR.
The log of observations is given in Table \ref{tab:obs}. Despite the different telescopes and instrumentation used in this study, the observation strategy was the same for all the runs, following the standard procedure of taking dithered frames of the target to subtract the sky. The field size was 2$^{\prime}{\times}$2$^{\prime}$ for the first three runs, for which a 256$^2$ pixel NICMOS3 array was employed. When using the 1024$^2$ pixels camera SOFI at the ESO-NTT telescope (run \#4), the field of view was
5$^{\prime}{\times}$5$^{\prime}$. 
In Table \ref{tab:obs}, columns record the telescopes and instruments used in each run, the total duration of the run, and the number of measurements for each cluster (last column: no. meas.). Note that, despite its five-night length, only a single point for the two clusters was obtained in run \#4, as this was a survey program on LMC clusters (see. \cite{Mucciarelli06}).

\begin{table*}[htbp]
\begin{tabular}{cccccccc}
Run Id. & Dates & Site & Tel. & Instr. & pix.sz &  field size & no. meas.\tabularnewline
\hline
1 & 02-08 Dec 1992 & LCO & Swope 1m & IR camera	& 0.53$^{\prime\prime}$ & 136$^{\prime\prime} \times$ 136$^{\prime\prime}$ & 5 \tabularnewline
2 & 12-16 Jan 1993 & ESO & MPI 2.2m & IRAC2b & 0.49$^{\prime\prime}$ & 125$^{\prime\prime} \times$ 125$^{\prime\prime}$ & 3 \tabularnewline
3 & 01-08 Nov 1993 & LCO & Swope 1m & IR camera & 0.53$^{\prime\prime}$ & 136$^{\prime\prime} \times$ 136$^{\prime\prime}$ & 5 \tabularnewline
4 & 27-31 Dec 2001 & ESO & NTT & SOFI & 0.29$^{\prime\prime}$ & 295$^{\prime\prime} \times$ 295$^{\prime\prime}$ & 1 \tabularnewline
\hline
\end{tabular} 
\caption{Observing Log}
\label{tab:obs}
\end{table*}

\subsection{Reductions}
In order to check the relative consistency of all the data-sets, two widely used packages have been used to generate object lists and magnitudes,  namely {\textsc {\small DAOPHOT}} \citep{DAOPHOT} and {\textsc {\small D{\scriptsize O}PHOT}} \citep{DoPHOT}. The detailed reduction strategy was slightly different for the various runs. For runs \#1 and \#2, sky flat-fields were taken during evening twilight; for run \#3, dome-flats were taken using the ESO-MPI 2.2m telescope dome; and for run \#4, dome-flats taken in the ESO-NTT dome were used together with screen-corrected images, as recommended in the SOFI manual. For all runs, dithered images were observed adopting off-source skies, taken some arcminutes away from the cluster center, and subtracting the sky frames to the single cluster images, before reconstructing the final scientific images by re-aligning the dithered source images. 

Bad pixel masks were used, whenever possible, by taking advantage of the dithering technique to remove bad pixels from the final images.  In practice, this technique did not always work very well, especially for the LCO runs, when the chip had a fairly large number of dead pixels. However, these artifacts yield measurements that were easily identifiable from their large photometric errors.

Photometry was performed with two different packages, in order to check for systematic differences due to the software, a strategy adopted in the HST Cepheid distance scale project\cite{Freedman01}. It should be noted that these two packages differ fundamentally in approach: {\textsc DAOPHOT} carries out multi-profile PSF fitting on a group of objects whose PSFs interact with each other, while {\textsc {\small D{\scriptsize O}PHOT}} adopts a single-profile fitting starting from the brightest source of the list, then subtracting the measured object from the image and proceeding in order of decreasing brightness.

When running \textsc{DAOPHOT} for all runs, all nights and all groups of three-filters (J, H, K$_S$) observations, a master list of candidates was built by carefully aligning and summing the three images, then running \textit{daofind} on the summed image. When running \textsc{\small D{\scriptsize O}PHOT}, the program was applied to the single images and the output lists cross-checked {\textit {a-posteriori}}. However, within the internal fitting error, measurements were fully consistent for high-S/N sources. For fainter sources, differences  were found to be somewhat larger, but always within 
3$\sigma$ of the (larger) internal error. Calibration errors and aperture correction errors have been added to the internal errors to obtain the total photometric errors.

\subsection{Calibrations}
During runs \#1, \#2 and \#3, standard stars were observed to obtain a primary calibration of the data. However, although the nights appeared to be photometric, a careful analysis of the calibrating standards revealed some inconsistencies that prevented us from obtaining  a well calibrated sample.
For this reason a further set of observations was taken during run \#4, when conditions were photometric. 
In order to check  the calibration, we cross-checked the output magnitudes against $\approx$ 500 stars from the 2MASS point-source catalog and with the photometry of \cite{Mucciarelli06}, based on the same data, and  found the values to be consistent in both cases. 
The other runs were then transformed to that photometry of run \#4 by applying a secondary calibration that, in all cases, consisted of a zero point term and a negligible color term.
The magnitudes were then transformed to the LCO system, for consistency with \cite{Persson04}. 
The relations used were obtained from \cite{Carpenter01}:

$$ \mathrm{J_{LCO} = J_{2MASS} + 0.020 - 0.014 \times (J_{2MASS}-K_{2MASS})} $$
$$ \mathrm{H_{LCO} = H_{2MASS} + 0.020 - 0.015 \times (H_{2MASS}-K_{2MASS})} $$
$$ \mathrm{K_{LCO} = K_{2MASS} + 0.015} $$


\section{The Photometric Sample \label{sect:phot} }

\begin{figure}[ht]
	\centering
	\includegraphics[width=8.5truecm]{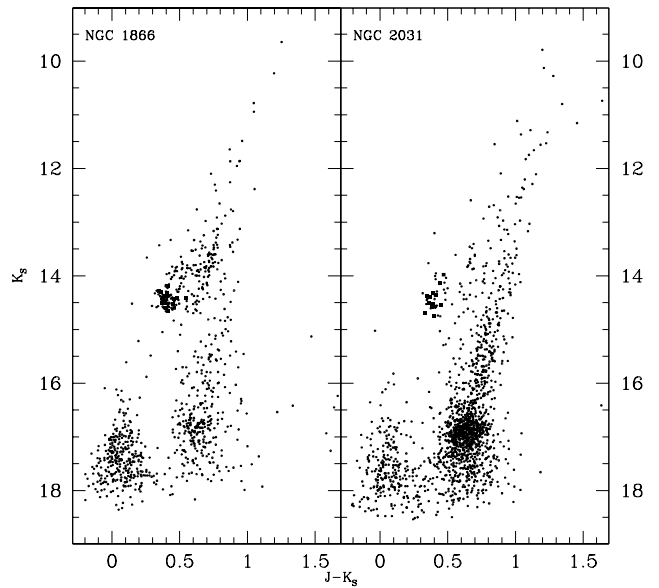}
	\caption{Near-IR color-magnitude diagrams for the two LMC clusters. Program Cepheids are identified by the filled squares.}
	\label{fig:cmd}
\end{figure}

Photometry lists for all runs and all nights were cross-matched in order to obtain: (1) a master list of the photometry; and (2) for each object, the list of measurements in all the images. In total, we obtained 14 photometric points, that were then averaged with a $\sigma$-clipping rejection algorithm to obtain the final photometry list of each cluster.
Cluster Cepheids were then identified and their magnitudes extracted to construct the light curves. Image coordinates were transformed onto the system of run \#4, for which an astrometric calibration was obtained from a subset of objects in the 2MASS catalog in common with those from our own catalog. The IR color-magnitude diagrams (CMDs) of NGC~1866 and NGC~2031 have been presented and discussed elsewhere \citep{Mucciarelli06}, although a preliminary version was published  in \cite{Maceroni02}, in order to identify the candidate TP-AGB stars for a spectroscopic study presented in that paper. Here, for the sake of completeness, we show the CMDs for the two clusters in Fig. \ref{fig:cmd}, in which our program Cepheids are identified by the filled squares.

\subsection{The Cepheids}

The Cepheids for NGC~1866 were selected from the lists of \cite{Welch91} and \cite{Welch93}, using the finding charts in those papers. For NGC~2031, the sample was obtained from \cite{Mateo92} and from our own unpublished lists. Since our frames covered the central 2$^{\prime}\times$2$^{\prime}$ of the cluster (with the exception of the SOFI field), the outermost Cepheids of NGC~1866 were not included in our IR list, but we identified all the variables in the cluster core. Of the 15 Cepheids of our sample, 8 are confirmed radial velocity members \citep[see][]{Welch91}. 
The fact that the outermost variables of NGC 1866 are not part of the presented sample prevents comparisons with some of the best-studied Cepheids, like the ones in \cite{Storm05} --- with which we have only a single object in common (HV 12202) --- but the inclusion of the inner variables increases considerably the overall sample. We compared the J and K magnitudes of HV12202 with the values reported in \cite{Storm05},  finding a good agreement. Hence, rather than using single point photometry  from run \#4 for these objects, we preferred to use the more precise mean values published by \cite{Storm05}. 

Because NGC~2031 is more compact, all the known objects fall within our fields with the exception of variables V5 and V7 that are not present in the current sample.  Their periods are, however, less precisely determined than the ones for NGC~1866. Indeed, as can be seen in Table \ref{tab:cep2031}, the periods have a typical precision of 0.01d, which likely explains the overall poorer quality of NGC 2031 light curves.

\begin{table}[ht]
{\scriptsize
\begin{tabular}{cccccccc}
Name & P (days) & $<$J$>$ & $\sigma$J & $<$H$>$ & $\sigma$H & $<$K$>$ & $\sigma$K \tabularnewline
\hline
 V4       & 3.3157  &  14.77 & 0.05 & 14.51 & 0.03  &  14.37 & 0.04 \tabularnewline
 We8      & 3.043   &  14.86 & 0.07 & 14.67 & 0.05  &  14.51 & 0.06 \tabularnewline
 V6 (ov)  & 1.9433  &  14.97 & 0.04 & 14.74 & 0.05  &  14.60 & 0.05 \tabularnewline
 V8 (ov)  & 2.0088  &  14.93 & 0.06 & 14.68 & 0.03  &  14.55 & 0.04 \tabularnewline
 V7       & 3.453   &  14.70 & 0.06 & 14.44 & 0.04  &  14.31 & 0.04 \tabularnewline
 HV12200  & 2.7248  &  14.96 & 0.08 & 14.71 & 0.06  &  14.53 & 0.09 \tabularnewline
 HV12202  & 3.10112 &  14.80 & 0.09 & 14.60 & 0.06  &  14.44 & 0.04 \tabularnewline
 We2      & 3.054   &  14.87 & 0.09 & 14.58 & 0.09  &  14.41 & 0.08 \tabularnewline
 We3      & 3.045   &  14.78 & 0.06 & 14.56 & 0.08  &  14.42 & 0.10 \tabularnewline
 We4      & 2.8604  &  14.83 & 0.13 & 14.58 & 0.13  &  14.46 & 0.12 \tabularnewline
 We5      & 3.1763  &  14.75 & 0.08 & 14.53 & 0.05  &  14.38 & 0.10 \tabularnewline
 We6      & 3.289   &  14.76 & 0.04 & 14.52 & 0.04  &  14.37 & 0.04 \tabularnewline
 WS5      & 2.895   &  14.87 & 0.06 & 14.61 & 0.07  &  14.50 & 0.07 \tabularnewline
 WS9      & 3.071   &  14.64 & 0.09 & 14.49 & 0.10  &  14.30 & 0.09 \tabularnewline
 WS11     & 3.0544  &  14.95 & 0.10 & 14.62 & 0.17  & 14.42  & 0.16 \tabularnewline
\hline
\end{tabular}  }
\caption{Mean IR magnitudes for Cepheids in NGC~1866.}
\label{tab:cep1866}
\end{table} 

\begin{table}[ht]
{\scriptsize
\begin{tabular}{cccccccc}
Name & P (days) & $<$J$>$ & $\sigma$J & $<$H$>$ & $\sigma$H & $<$K$>$ & $\sigma$K \tabularnewline
\hline
 V1      &    3.07  & 14.71 & 0.08 &  14.38 & 0.07 &  14.38 & 0.06 \tabularnewline
 V2      &    4.43  & 14.43 & 0.08 &  14.03 & 0.05 &  13.98 & 0.05 \tabularnewline
 V3      &    3.96  & 14.56 & 0.09 &  14.16 & 0.08 &  14.14 & 0.09 \tabularnewline
 V4      &    3.43  & 14.73 & 0.07 &  14.37 & 0.09 &  14.39 & 0.05 \tabularnewline
 V5      &    3.32  & 14.68 & 0.12 &  14.32 & 0.12 &  14.31 & 0.12 \tabularnewline
 V6      &    3.03  & 14.85 & 0.14 &  14.50 & 0.08 &  14.51 & 0.09 \tabularnewline
 V7      &    3.13  & 14.94 & 0.08 &  14.56 & 0.08 &  14.58 & 0.09 \tabularnewline
 V8      &    3.27  & 14.75 & 0.08 &  14.39 & 0.08 &  14.36 & 0.08 \tabularnewline
 V9      &    2.95  & 14.89 & 0.07 &  14.53 & 0.08 &  14.52 & 0.06 \tabularnewline
 V11     &    2.82  & 15.01 & 0.12 &  14.67 & 0.07 &  14.70 & 0.07 \tabularnewline
 V12 (ov)&    1.84  & 15.12 & 0.08 &  14.73 & 0.09 &  14.74 & 0.07 \tabularnewline  
 V13     &    3.20  & 14.75 & 0.11 &  14.47 & 0.11 &  14.40 & 0.08 \tabularnewline
 V14     &    2.97  & 14.97 & 0.05 &  14.56 & 0.06 &  14.54 & 0.06 \tabularnewline

\hline
\end{tabular}  }
\caption{Mean IR magnitudes for Cepheids in NGC~2031.}
\label{tab:cep2031}
\end{table}

\section{Light Curves and PL Relations \label{sect:pl} }

\subsection{Light Curves and Mean Magnitudes}

The adopted periods for NGC~1866 are taken from \cite{Welch91} and \cite{Welch93}. For a few Cepheids, periods have been re-determined by \cite{Musella06} and show no sign of significant changes between the two epochs. The periods of the NGC~2031 Cepheids have been taken from \cite{Mateo92} and from our own list. We folded the data using the adopted periods obtaining light curves in all the three bands  and the color curves in $\mathrm{J-K}$. Light  and color curves are shown in Fig. \ref{fig:lc1866:first} for NGC 1866, and in Fig. \ref{subfig:continued:second} for NGC 2031. 
No shift has been applied to the phases in order to have the zero of the phase as first point, as is usually done, since this has been  found to be unimportant for the determination of mean magnitudes. Note that variables V5 and V7 in NGC 2031 have only one measurement and do not appear in the light curve plots.
As can be seen in the figures, the sampling is far from optimal and, in some cases, the scatter is relatively large. Nevertheless, the periodic shape is almost always clearly defined, with the exception of a few outliers that usually have larger magnitude errors. For these, we checked the images to verify the presence of glitches or problems and we found that always, in these cases, there was either a problem of crowding, especially in the J filter, combined with a larger FWHM and/or patches of cosmic rays or dead pixels that affected the measurement. In fact, the points that are more scattered around the fiducial light curve are invariably from runs \#1 and \#3, when the Las Campanas 1m IR camera had poor cosmetic properties (namely a large patch of bad pixels in the right part of the chip). The points with the largest errors have been removed while we kept the others for the determination of the mean megnitude.

As expected, the global amplitude of the oscillations is of the order of a few tenths of magnitude. We calculated mean magnitudes by averaging the data, rather than fitting a template to the light curves. This method is preferred for its higher speed with respect to the classical, more accurate, template curve fitting, because the amplitudes of the Cepheids' light curves are relatively small and our data provide a sparse but relatively uniform phase coverage. The different uncertainties from variable to variable are mostly due to the scatter along the light curve and to individual nonuniformities in period sampling. Objects having light curves with the poorest sampling suffer from a larger uncertainty in the mean magnitude. Moreover, since the light curve sampling is sparse, fitting to a template light curve does not reduce significantly the error on the mean magnitude. In order to reduce the effect of outliers and obtain a robust estimate, an algorithm based on a biweight location estimate has been used. The biweight location is a robust estimator of the ``mean" which uses weights defined as a function of the deviation from the median and calculated in an iterative way. In this context, robustness means that its sensitivity to single outliers is strongly reduced with respect to classic estimators. For a detailed description of the method \citep[see, e.g.][]{Mosteller77}. 
The average error on the mean magnitude is of the order of the error on the single measurement and slightly larger than that expected from a template fitting of a well sampled light curve.  Table \ref{tab:cep1866} records mean magnitudes in the three bands for 15 Cepheids of NGC~1866. Data for the 13 Cepheids in NGC~2031 are given in Table \ref{tab:cep2031}. 



\subsection{Period-Luminosity Relation }

In order to construct PL relations in all the near-IR bands, we first need to adopt values for the cluster reddenings.
The case of NGC~1866 has been widely discussed in the past. A value of $\mathrm{E(B-V)} = 0.12$~mag has recently been proposed in \cite{Groenewegen03} from the Wesenheit function, but this value was questioned by \cite{Storm05} based on an IR surface brightness analysis. We therefore use the canonical value of $\mathrm{E(B-V)} = 0.06$~mag for this cluster \citep[see][ and references therein]{Storm05}.
It should be pointed out that, by adopting $\mathrm{A_J} = 0.9546 \times \mathrm{E(B-V)}$, $\mathrm{A_H} = 0.6110 \times \mathrm{E(B-V)}$, $\mathrm{A_{K_S}} = 0.3842 \times \mathrm{E(B-V)}$ \citep[see][]{CCM89}, the difference in using one value or the other values is of the order of 0.04~mag in $\mathrm{J}$, and 0.02~mag in $\mathrm{K_S}$. 
By contrast, NGC~2031 lies in a crowded region close to the LMC bar, so its reddening is correspondingly higher. 
Here we adopt a value of \cite{Mould93} $\mathrm{E(B-V)} = 0.18$~mag, leading to correction terms of: $\mathrm{A_J} = 0.17$~mag, $\mathrm{A_H} = 0.11$~mag, $\mathrm{A_{K_S}} = 0.07$~mag.

Given the small range of periods covered by our program Cepheids, a fit of the PL relation to these objects only is of limited  use, due to the large uncertainties on the derived coefficients. As an example, the PL fit in the K band for NGC~1886 gives: $\mathrm{M_K(1866) = -2.79\pm 0.31 * log(\mathrm{P}) + 15.78\pm0.15}$.

Our sample can instead be used for a comparison with field Cepheids, namely the  sample of \cite{Persson04} in which Cepheids with period shorter than $\sim$3 days are absent. This comparison has two goals: (1) to check if the  cluster and field Cepheids obey the same relation and, if so, to derive an overall PL relation extending over an enhanced period range; and (2) to check if the break in the slope of the PL relation at about $\mathrm{P = 10 d}$, discussed by \cite{Ngeow05,Ngeow06} is present in our sample.
Figure \ref{fig:allper} shows PL relations for the sample discussed in this work, the \cite{Storm05} variables not included in our sample, and the sample of \cite{Persson04}.  Cluster Cepheids appear in the plot, as expected, as a extension of the sequence of \cite{Persson04} and seem to describe the same PL relation, as was previously noted by Storm et al. (2005) (see their Fig.~15). 

Figure \ref{fig:shortper} is a magnification of the previous figure at $\mathrm{log(P)} < 1$ which shows this fact more clearly. Note that two of the variables in the NGC~2031 sample have longer period than the bulk of the cluster Cepheids and are likely field variables. If confirmed, this would provide further evidence that field and cluster Cepheids describe the same relation.

In addition, \cite{Keller06} report a mean metallicity of $\mathrm{[Fe/H] = -0.34 \pm 0.03}$ for their field sample, which is quite close to the metallicities of our two target clusters. Of course, the effect of metallicity on the PL relation --- if such a dependence does indeed exist --- is greatly reduced in the near-IR, especially in the K band. Thus, the fact that our cluster Cepheids extend the tight relation of \cite{Persson04} toward short periods is perhaps not surprising. 

A new fit of the PL relation using the entire sample has been performed in each of the three IR bands, and the results compared with \cite{Persson04}. The output is shown in Fig. \ref{fig:shortper}  as the red dashed line. The derived coefficients (see Fig. \ref{fig:allper}) are essentially identical to the ones found by \cite{Persson04}. 

There is an ongoing debate over possible non-linearities in the PL relation of LMC Cepheids. \cite{Ngeow05,Ngeow06} and \cite{Sandage04} reported that the PL relation of the LMC Cepheids shows evidence for a change in slope at about $P = 10$ days. This effect is most apparent in the optical (i.e, V-, R- and I-band) PL relations, while \cite{Ngeow05,Ngeow06} show that, at least in the J and H bands, it is present also in the IR. On the other hand, \cite{Gieren05}, by using IR surface brightness technique, found a {\it constant} slope, in agreement with the trend for Galactic Cepheids. From Figures \ref{fig:allper} and \ref{fig:shortper}, it can be seen that the slope looks essentially constant in all the three IR bands. In fact, by adding the short-period cluster Cepheids to the Persson sample, the slope is virtually unchanged, if not slightly flatter than the original Persson fit. Hence, the new fit with the extended period range seems to confirm the hypothesis of constant slope in the LMC Cepheids.

\begin{landscape}
\begin{figure}[ht]
   \centering
   \hspace{-5.0cm}
  \vbox{
   \vspace{-5.1cm}
   \subfloat{
	\label{subfig:continued:first}
	\includegraphics[width=1.0\textwidth]{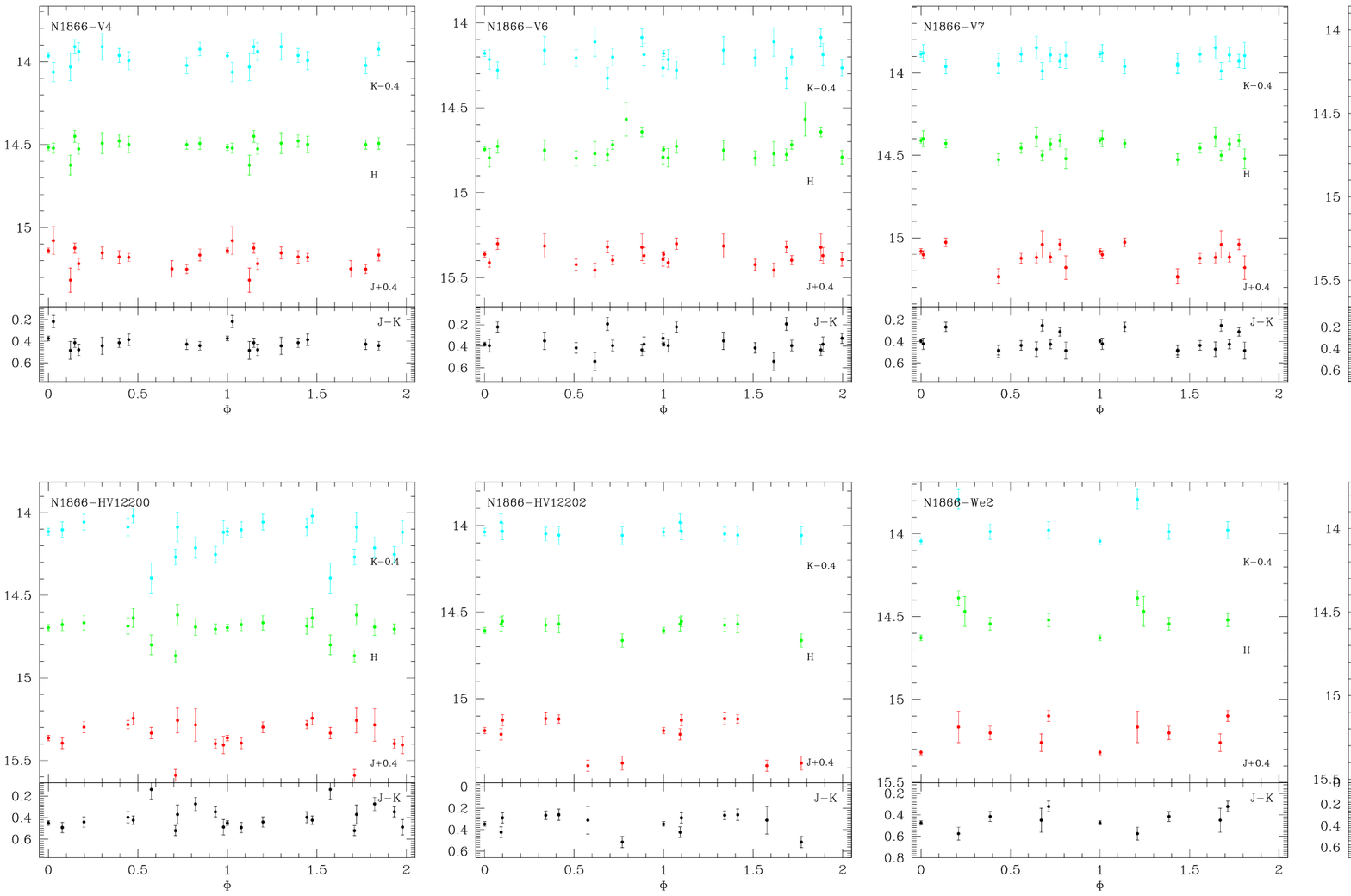}
   }
   \vspace{-5.0cm}
   \caption{Light curves for Cepheids in NGC~1866.}
   \label{fig:lc1866:first}
  }
\end{figure}
\end{landscape}

\begin{landscape}
\begin{figure}[ht]
   \ContinuedFloat
   \hspace{-5.0cm}
   \centering
  \vbox{
   \vspace{-5.1cm}
   \subfloat[{\it cont.}]{
	\label{subfig:continued:second}
	\includegraphics[width=1.0\textwidth]{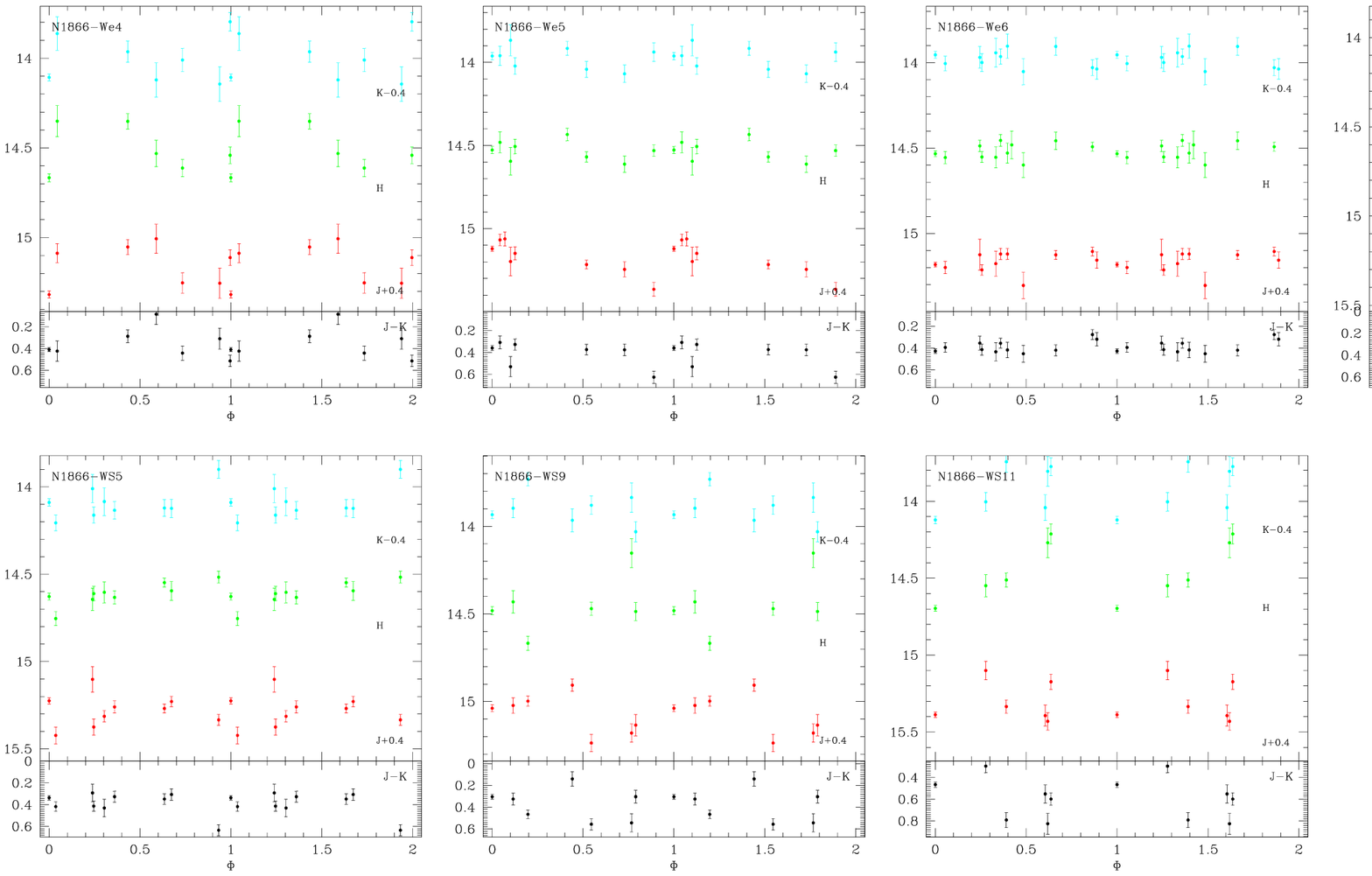}
   }
   \vspace{-6.0cm}
   \caption{{\it continued}}
  }
   \label{fig:lc1866:second}
\end{figure}
\end{landscape}

\begin{landscape}
\begin{figure}[ht]
   \centering
  \vbox{
   \vspace{-5.1cm}
   \subfloat{
	\label{subfig:cep2031:first}
	\includegraphics[width=1.0\textwidth]{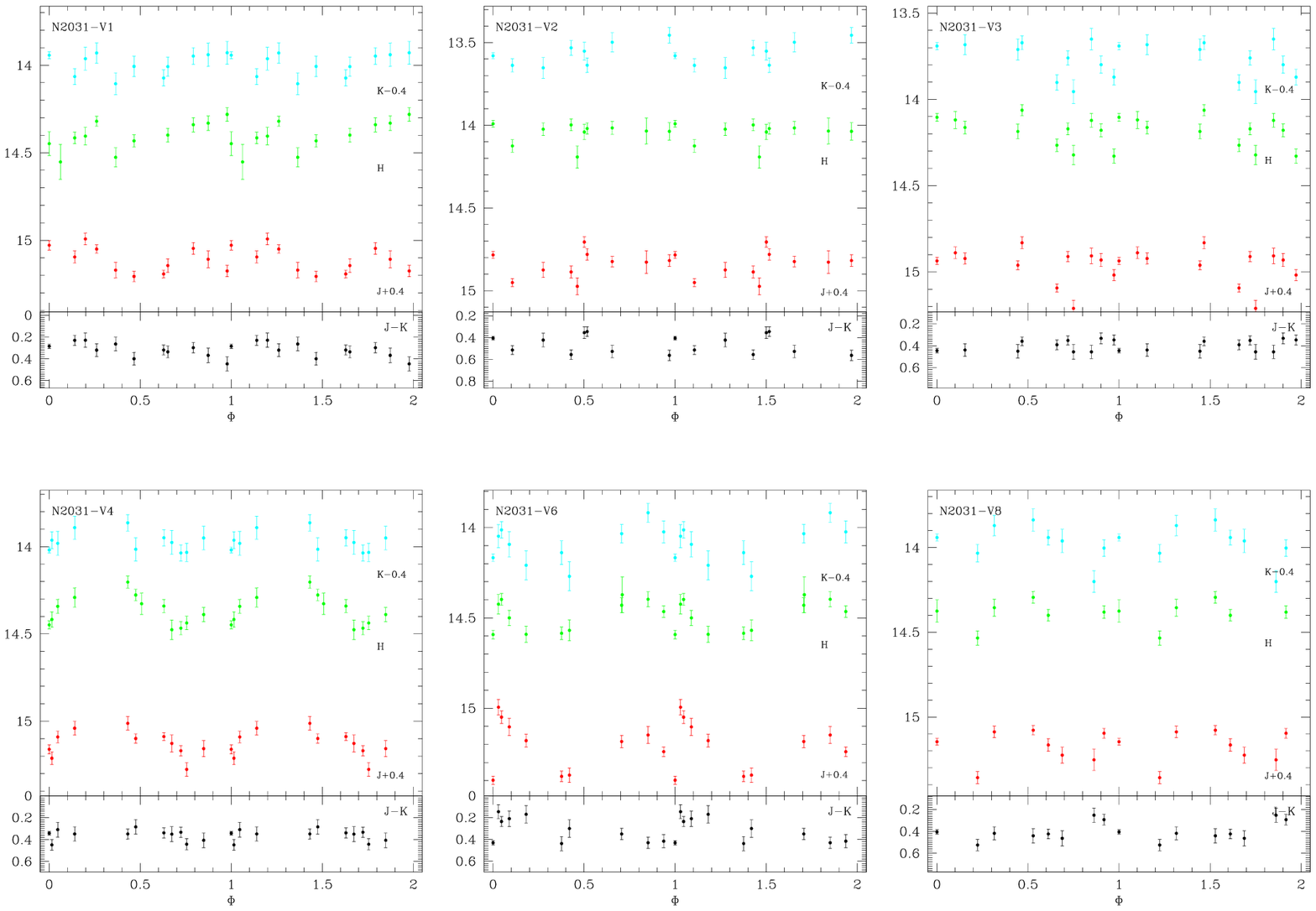}
   }
   \vspace{-5.0cm}
   \caption{Light curves for Cepheids in NGC~2031.}
   \label{fig:lc2031:first}
  }
\end{figure}
\end{landscape}

\begin{landscape}
\begin{figure}[ht]
   \ContinuedFloat
   \centering
  \vbox{
   \vspace{-5.1cm}
   \subfloat[{\it continued}]{
	\label{subfig:cep2031:second}
	\includegraphics[width=1.0\textwidth]{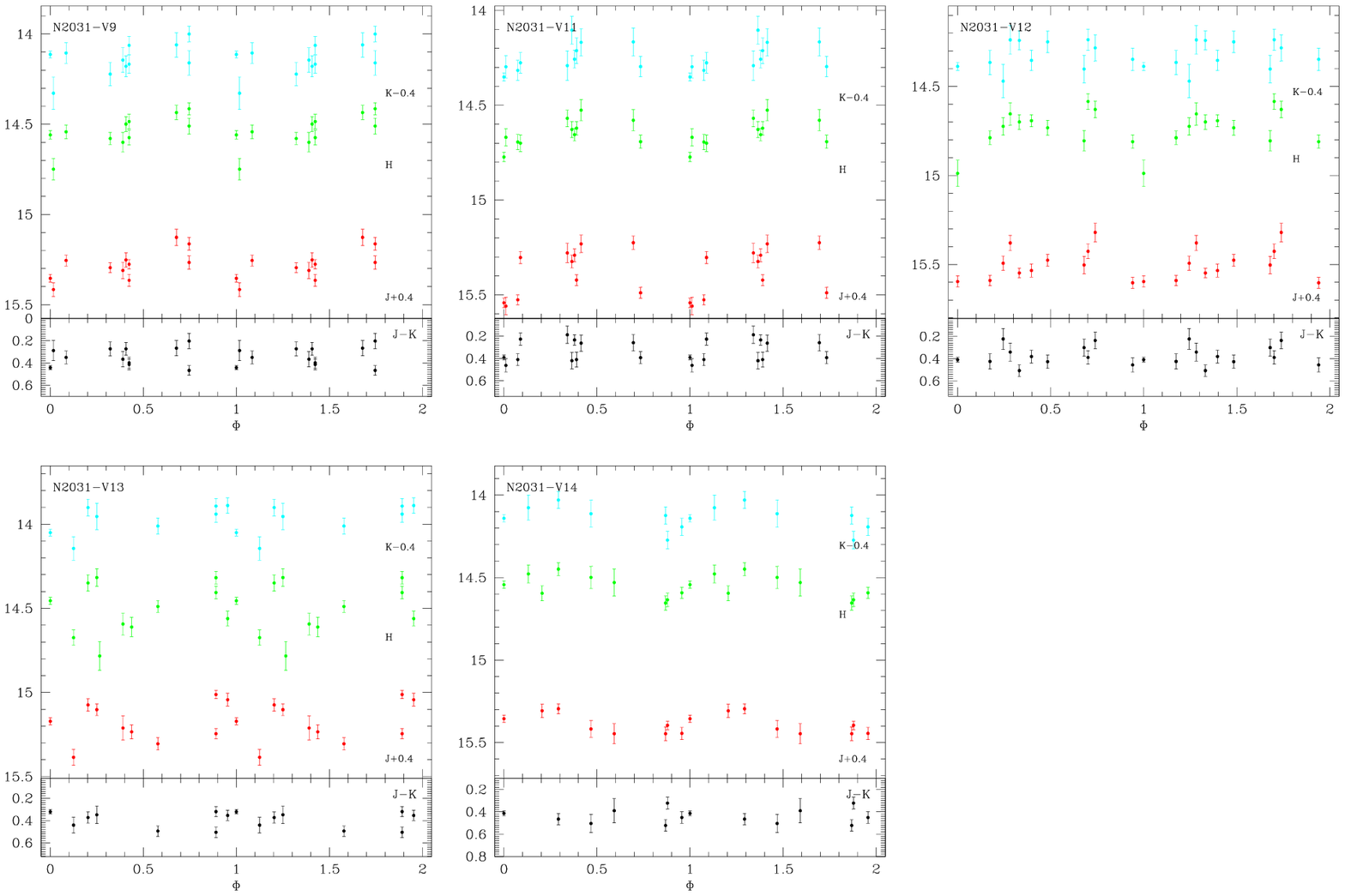}
   }
   \vspace{-6.0cm}
   \caption{{\it continued}}
   \label{fig:lc2031:second}
  }
\end{figure}
\end{landscape}

We have also made a comparison with the predicted PL relations from nonlinear convective pulsation models, for metallicities and period ranges appropriate for our sample  
\citep[see][ and references therein]{Brocato04,Bono99b,Bono02}. The theoretical PL relations in J and K have similar values for the slope,
$$M_J=-3.00\pm{0.04} \log{P} -2.40\pm{0.06}$$
$$M_K=-3.20\pm{0.05} \log{P} -2.66\pm{0.05},$$
with intrinsic dispersions of 0.15~mag and 0.13~mag, respectively.
Comparison of theoretical and observed relations yields an estimate for the DM of LMC: $\mathrm{(m-M)_0 = 18.63 \pm 0.11}$~mag for J and $\mathrm{(m-M)_0 = 18.62 \pm 0.10}$~mag for K, that is slightly higher than the most recent estimates for the DM of the LMC obtained from comparison with theoretical models \citep[18.54~mag ][]{Marconi06,Keller06} or with observational methods \citep[18.52~mag][]{Clementini03}, \citep[18.56~mag][]{Gieren05}. In particular, \cite{Alves04} has reviewed a series of distance measurements, finding an average DM of 18.50 $\pm$ 0.02~mag.

\begin{figure}[ht]
 \centering
 \hspace{-0.5cm}
  \includegraphics[width=0.5\textwidth]{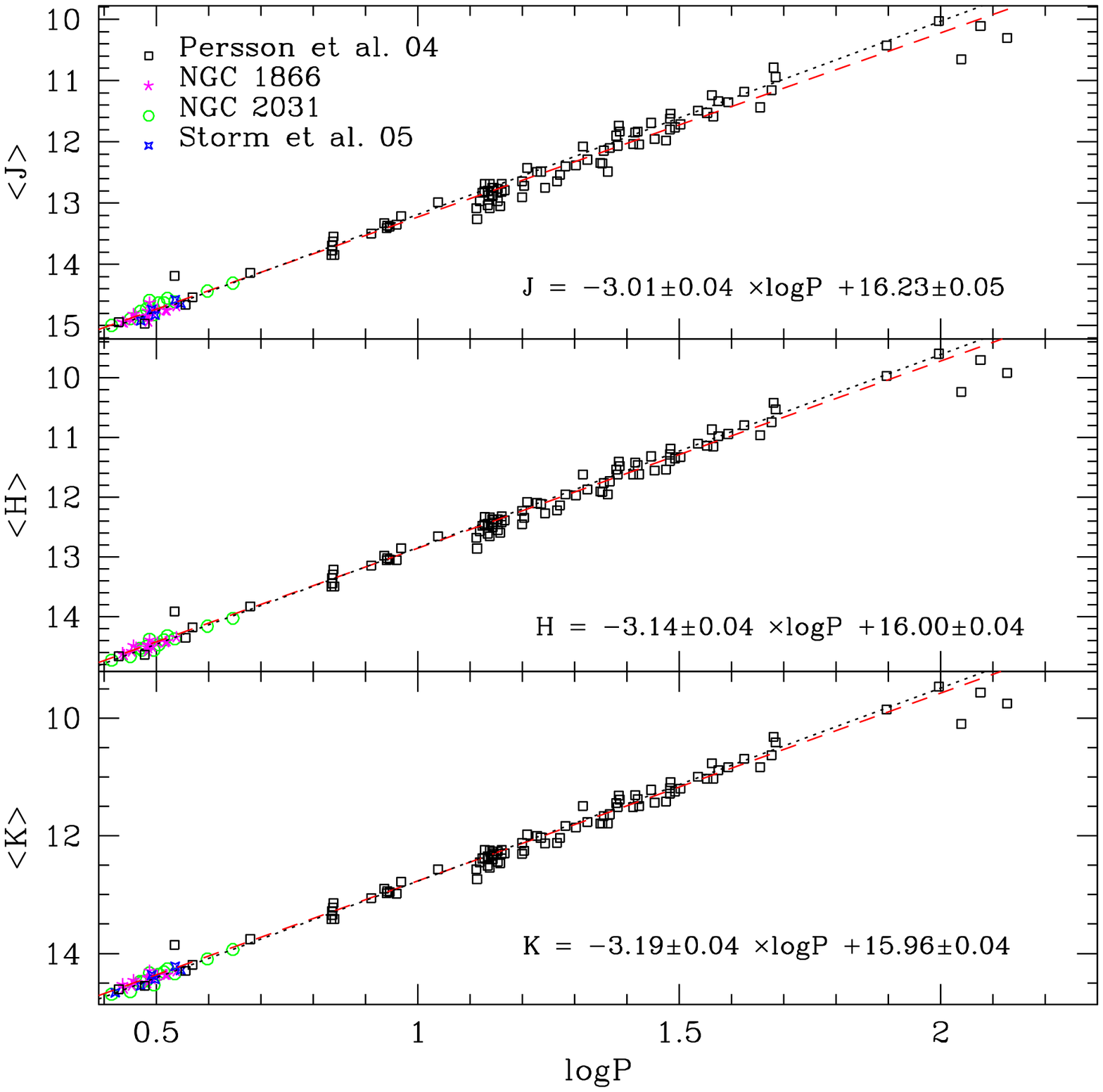}
 \caption{Overall PL relations for the field Cepheids of \cite{Persson04} and the two target clusters. The black dotted line represents the PL relation of \cite{Persson04}, the red dashed line shows the overall fit obtained with the combined cluster sample and the sample of \cite{Persson04}. The symbols for Cepheids are given in the plot.}
 \label{fig:allper}
\end{figure}

The zeropoints of theoretical PL relations depend on both the physical and numerical assumptions required for the construction of the pulsation models. If over the same period range, mildly overluminous pulsation models \citep[see][]{Bono99b,Brocato04} are adopted instead of the canonical ones (following recent suggestions that Cepheid masses should be smaller than the values predicted on the basis of the canonical evolutionary scenario; see, e.g., Brocato et al. 2004,Caputo et al. 2005), then the theoretical fundamental PL relations become
$$M_J=-3.11\pm{0.05} \log{P} -2.05\pm{0.07}$$
$$M_K=-3.27\pm{0.04} \log{P} -2.37\pm{0.05},$$
with an instrinsic dispersion of 0.19~mag and 0.15~mag.
By applying these relations to the fundamental Cepheid data for NGC1866 we obtain the mean apparent distance moduli  $\mathrm{(m-M)_0 = 18.45 \pm 0.11}$~mag in J and $\mathrm{(m-M)_0 = 18.42 \pm 0.10}$~mag for K, in better agreement with the short distance scale.
Note that these are random errors that include all the observational and theoretical source of random uncertainty but do not take into account possible systematics due to physics used in the models. These include uncertainties arising from the treatment of convection (i.e., the mixing length parameter), the mass-luminosity relation and the atmospheric models used to obtain observed values from theoretical luminosity and temperature. The effect of the equation of state is negligible, and varying the opacity does not produce 
effects on the models, provided that one uses the most recent compilations (\citep[see][ for a discussion of this dependence]{Petroni03}. The effects of mixing length is under investigation (Marconi et al., in preparation) and preliminary results indicate that by increasing $\alpha$ from 1.5 to 1.8, the PL relation slope becomes slightly steeper. Such a change is expected to have a relatively minor impact on the PL relations derived above.

\begin{figure}[ht]
 \includegraphics[width=0.5\textwidth]{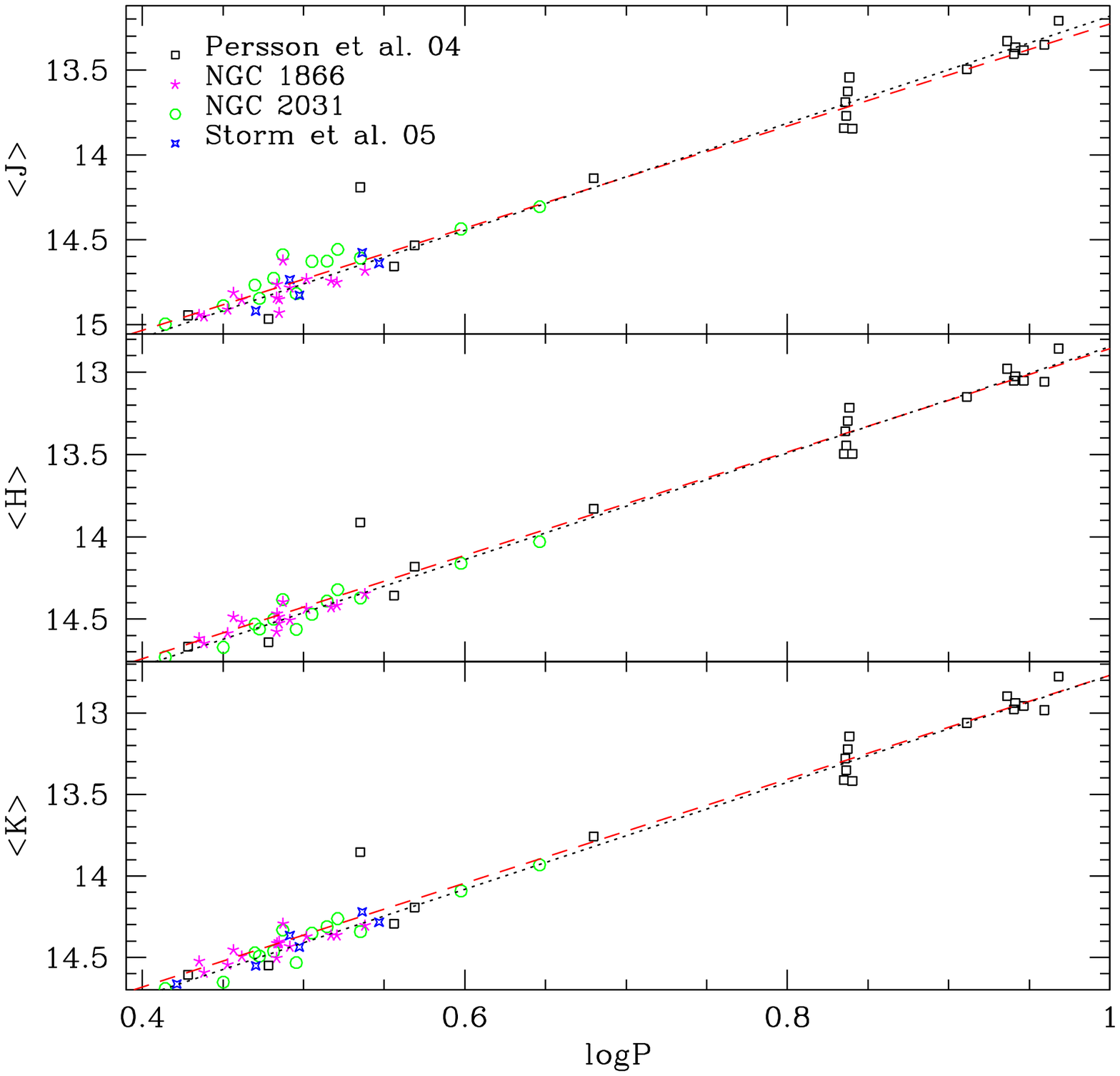}
 \caption{Magnified view of the PL relation for log(P) $<$ 1. Lines and symbols are the same as in Fig. \ref{fig:allper}.}
 \label{fig:shortper}
\end{figure}

\section{Discussion and Summary \label{sect:summ} }

We have presented IR light and color curves for 15  and 13 Cepheids in the young LMC clusters NGC~1866 and NGC~2031, respectively. The light curves are of sufficient quality to derive mean J, H, and K$_{\mathrm{S}}$ magnitudes, allowing us to compare our results on cluster Cepheids with the analogous study done on field Cepheids by \cite{Persson04}, with theoretical models by \cite{Bono99b,Brocato04} and, for NGC 1866, with previous studies from the literature. 

The good agreement we find with field Cepheids has allowed us to extend the PL relation of \cite{Persson04} to shorter periods. A redetermination of the PL relation yields slopes that are in agreement with recent theoretical models. To obtain agreement with recent determinations of the LMC DM that give the same value (i.e., $\mathrm{(m-M)_0 = 18.54}$) for both RR Lyrae \citep{Marconi06} and with Cepheids \citep{Keller06} would require overluminous models. In this case, whether the overluminosity is due to convective core overshooting or mass-loss is still a point of active debate. 

A comparison of results obtained on NGC~1866 by different authors and different methods shows that some authors favour canonical models with the presence of binaries, while others conclude a certain amount of core overshooting is required. For example, \cite{Walker01} use MS fitting on HST data to derive a DM of $\mathrm{(m-M)_0 = 18.35\pm0.06}$
by using the observed Hyades sequence and theoretical models to take into account metallicity differences. Ground-based studies by \cite{Testa99} and \cite{Barmina02} used essentially the same data but yielded the opposite conclusion about the overshooting issue. The only point of accordance seem to be the presence of binaries at the tip of the MS. A later redetermination by \cite{Salaris03} seems to confirm the DM of NGC~1866 found by \cite{Walker01} and finds that the surrounding field has a longer DM of $\mathrm{(m-M)_0 = 18.53\pm0.07}$ (i.e., presumably due to the tilt of the LMC intermediate-age disk). 

In this work, the PL relations for the two clusters give essentially the same results, indicating that the two clusters have nearly the same DM. The structure of the young LMC --- where by ``young" we mean the sample of Cepheids analyzed in \cite{Persson04} --- may be described as a tilted disk, \citep[see, e.g., the discussion in][]{Alves04} with the north-east quadrant closer to the Earth, and with some amount of thickness \citep{vanderMarel02}. The fact that the two clusters share essentially the same DM is consistent with NGC~1866 lying {\it behind} the disk and at the same distance of the bar where NGC~2031 is located. 
Although our data cannot solve this important dilemma, it is worth emphasizing some  outstanding issues that need to be resolved. Probably the most crucial point is the reddening --- a better determination of this important parameter is clearly needed. On the other hand, metallicity has a less dramatic effect, particularly at near-IR wavelengths.

New models have recently made some progress towards resolving the longstanding discrepancy between evolutionary and pulsational masses; in fact, the same models are applied to cluster and field Cepheids. Finally, since the largest uncertainties seem to arise from the current observational material, the ultimate solution to the LMC distance problem will almost certainly require increased samples that include Cepheids in additional LMC clusters.

\begin{acknowledgements}
It's a pleasure to thank Giuseppe Bono for many useful discussions on this paper. This publication makes use of data products from the Two Micron All Sky Survey, which is a joint project of the University of Massachusetts and the Infrared Processing and Analysis Center/California Institute of Technology, funded by the National Aeronautics and Space Administration and the National Science Foundation. We would like to thank the referee, Dr. W.P. Gieren, for his useful comments and suggestions that greatly  improved the paper.
\end{acknowledgements}

\bibliographystyle{aa}
\bibliography{paper_final}

\end{document}